# Nanostructured Plasmonic Metal Surfaces as Optical Components for Infrared Imaging and Sensing

Jyotirmoy Mandal,[1]* John Brewer,[1] Sagar Mandal,[2] Aaswath P. Raman[1]*

[1] Department of Materials Science and Engineering, University of California, Los Angeles, USA

[2] Independent researcher, Seattle, USA

Contact: Jyotirmoy Mandal (jyotirmoymandal@ucla.edu) and Aaswath P. Raman (aaswath@ucla.edu)

**Abstract**

*Thermal imaging and sensing technologies offer critical information about our thermally radiant world, and in recent years, have seen dramatic increases in usage for a range of applications. However, the cost and technical finesse of manufacturing infrared optical components remain a major barrier towards the democratization of these technologies. In this report, we present a solution-processed plasmonic reflective filter (PRF) as a scalable and low-cost thermal infrared optic. The PRF selectively absorbs sunlight and specularly reflects thermal infrared (TIR) wavelengths with performance comparable to state-of-the-art TIR optics made of materials like Germanium. Unlike traditional infrared optical components, however, the PRF can be conveniently fabricated using low-cost materials and a 'dip-and-dry' chemical synthesis technique, and crucially, has manufacturing costs that are orders of magnitude lower. We experimentally demonstrate the PRF's core optical functionality, as well as its integration into infrared imaging and sensing systems without compromising their thermographic or radiometric capabilities. From a practical standpoint, the low cost and convenient fabricability of the PRF represent a significant advance towards making the benefits of thermal imaging and sensing systems more affordable and accessible. Scientifically, our work demonstrates a previously unexplored optical functionality and a new direction for versatile chemical synthesis in designing optical components.*

**Introduction**

Infrared imaging, also known as thermography or 'heat vision', offers a radically different view of the world from what we see. Human eyes and traditional cameras sense radiation in the 0.4-0.7 µm wavelengths ($\lambda$), where light from the sun and artificial sources, and the high optical transmittance of the atmosphere, enable imaging of our environment. A similar phenomenon occurs in certain bands within the thermal infrared wavelengths (TIR, $\lambda \sim$ 2.5-50 µm), including the long-wavelength infrared band (LWIR, $\lambda \sim$ 8-13 µm), where the atmosphere transmits thermal radiation emanating from objects, allowing them to be imaged. Just as our vision informs us about textures and colors of objects, thermography provides information about their emittances and temperatures, offering an alternative way to optically distinguish objects, or in the absence of illumination, be the sole imaging platform.

Because thermography enables the detection, imaging and characterization of heat from our surroundings, it is used in diverse applications such as astronomy, atmospheric monitoring, defense, industrial quality control, thermal fault detection (e.g., in building insulation), heat sensing in autonomous vehicles and medical screening. The last of these applications was recently prominent in the context of the COVID-19 pandemic. However, state-of-the-art thermal cameras contain optical components and microbolometer sensors which are costly and require technical finesse to manufacture. Consequently, the use of thermography has historically been limited to high-end consumers. Making thermal imaging a more accessible and affordable technology could significantly impact applications in a wide range of industries, and lead to beneficial uses that are today limited due to the costs of conventional thermographic systems.

The limited accessibility and affordability of thermal imaging and sensing systems arises in large part from the material and fabrication costs of their optical components. Typically, these systems require the filtering of solar heat and direction of TIR onto a sensor, with an additional requirement of specular yield in the TIR for imaging systems. Traditionally, IR-imaging and sensing systems have achieved this using germanium-based transmission optics. A semiconductor with a band-gap of 0.66 eV, germanium (Ge) intrinsically absorbs and filters out noise in the solar wavelengths < 1.9 µm, while transmitting longer wavelengths up to 16 µm onto detectors. However, as a material, Ge is expensive, and fashioning its brittle form into precision optical components requires advanced manufacturing processes. Furthermore, the high infrared refractive index (~4) of Ge requires antireflection coatings on its surfaces to maximize TIR transmission. These factors make Ge-based optics comparatively difficult to manufacture and expensive to use. Ge-based optics also have operational limitations. For instance, its transmittance at λ~1.9-2.5 µm solar wavelengths[1,2] lets in noise. Its transmittance also falls and becomes non-monotonic midway through the LWIR at λ~11.5 µm[1,2] – which could complicate radiometric measurements. Furthermore, the increase in intrinsic carrier density with temperature makes Ge's transmittance and dispersion sensitive to temperature, further compounding radiometric measurements, and renders it opaque above 80˚C, preventing its use in high-temperature settings.[3,4]

Beyond Ge, recent advances in chalcogenide glasses and diffractive lenses are promising, however they are subject to one or more of the above limitations.[5–9] Alternatives like polyethene, zinc selenide (ZnSe) or silicon lenses, or gold-coated reflective surfaces have lower costs and wider transmittance or reflectance bands, but do not adequately filter sunlight. Thus, there remains considerable room for improvement in terms of cost, fabricability, and optical performance – with the first two being crucial for increasing the scope of, or access to, thermography. One way of achieving this may be through the use of non-traditional optical components in thermal imaging and sensing systems.

In this report, we present a solution-processed plasmonic reflective filter (PRF) as an optical component for use in thermal imaging and sensing systems. Our design consists of a layer of plasmonic metal nanoparticles deposited on a metal surface (Fig. 1A), which behaves as a specular reflector of incident TIR radiation (λ> 2.5 µm) while emitting little of its own, and strongly absorbing incident shortwave radiation (λ< 2.5 µm) (Fig. 1B). The spectrally selective, specular reflectance allows the PRF to direct thermal signatures from objects onto infrared sensors effectively, while filtering out shortwave noise (e.g. sunlight) (Fig. 1C and D). We first theoretically explore the optical mechanism that leads to the functionality of the PRF, and then demonstrate a PRF made using an exceptionally scalable solution-based 'dip-and-dry' technique, whose optical functionality is verified through spectrophotometry and Twyman-Green interferometry. We then show the integrability of the PRF into existing thermal imaging and sensing systems through radiometric characterizations and compare its performance relative to Germanium (Ge)-based designs used in today's infrared (IR) imaging and sensing systems. Our work shows that the PRF can yield a state-of-the-art, tunable, and versatile optical performance, but at material and fabrication costs that are lower by $10^2$-$10^3$ compared to that of Ge-based optics. From a practical standpoint, this may represent a significant advance in development of high-performance and low-cost substitute for Ge-based optics in thermal imagers and sensors. Scientifically, the PRF represents a novel IR optical functionality, and opens a new avenue to non-traditional fabrication pathways that combine optical design with chemical synthesis.

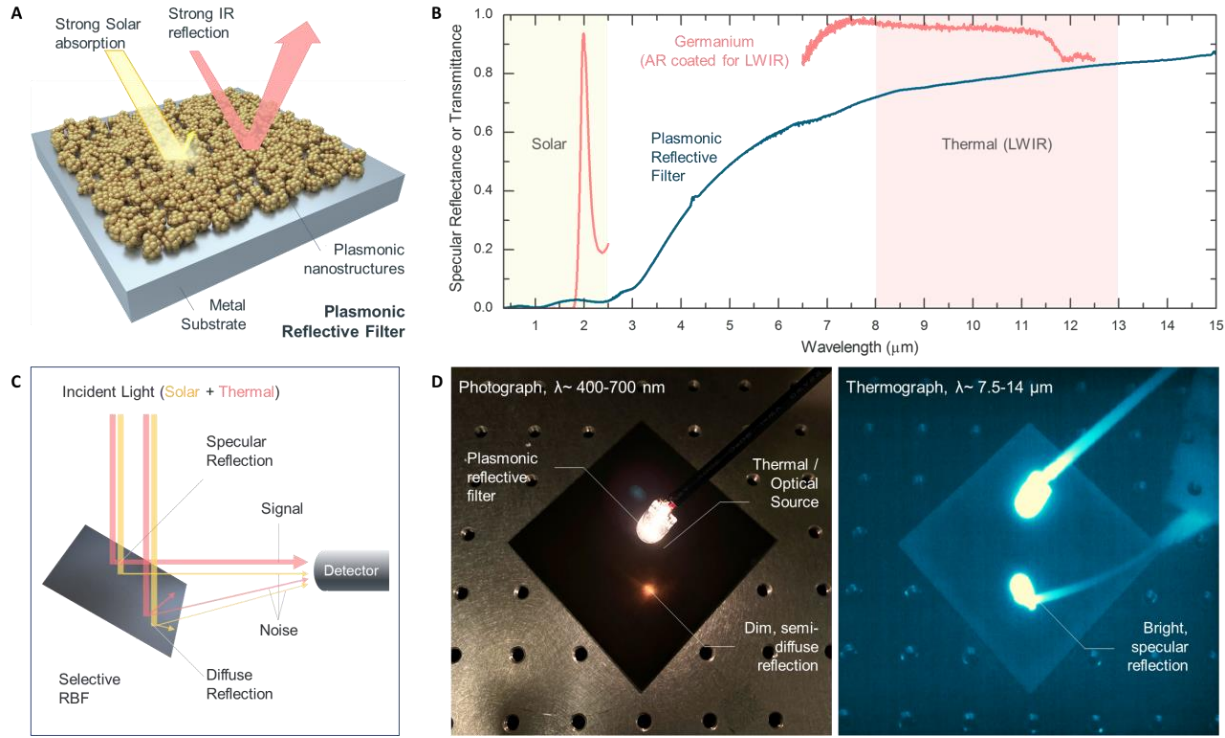

**Figure 1**. The plasmonic reflective filter (PRF) and its functionality. (**A**) Schematic of the PRF, showing a plasmonic nanostructure coated metal substrate that acts as a selective reflector of thermal infrared (TIR) radiation. (**B**) Experimentally measured specular reflectance of the PRF presented alongside that of an antireflection-coated (for LWIR) germanium window for comparison. As evident, the PRF shows very little shortwave (λ< 2.5 µm) specular reflection, and a high specular reflection across the TIR, and particularly long-wavelength infrared (LWIR, λ~8-13 µm) wavelengths where the atmosphere is transparent. (**C**) Consequently, the PRF can minimize incident noise in the form of sunlight, and direct thermal signatures from the environment onto imaging sensors or detectors, as evident from (**D**) photo- and thermographs of a tungsten lamp over a PRF.

## Theory

Optical components used in IR-imaging and sensing systems must direct incident TIR radiation onto a sensor, while filtering out sunlight which would otherwise be registered as noise. Typically, the detection occurs in the atmospheric transmission windows, such as the long-wavelength infrared band (LWIR, λ~ 8-13 µm), where thermal emissions from typical terrestrial objects peak, or the mid-wavelength infrared band (MWIR, λ~ 3-5 µm), which carries radiation from hot objects. An additional requirement for optics in imaging systems is that the TIR is directed without distortion of the incident wavefronts, which requires maximizing specularity.

We propose that a plasmonic nanoparticle-coated metal mirror could offer these optical functionalities in a reflective arrangement while overcoming the limitations of traditionally-used optical components in the infrared. Our approach is based on the well-known phenomenon that metal nanoparticles undergo plasmonic oscillations when exposed to solar wavelengths, resulting in strong scattering and absorption of light. This enables an ensemble of plasmonic metal nanoparticles to act as a efficient solar absorber.[10,11] The longer thermal wavelengths, however, interact weakly with individual nanoparticles due to their small cross sections, causing the ensemble to behave as a weakly-lossy or reflective effective medium. Thus, by backing a layer of plasmonic nanoparticles with an intrinsically IR-reflective homogenous metal layer, one should be able to create a selective absorptance in the solar wavelengths and reflectance in the TIR.

**Experimental Demonstration and Optical and Radiometric Characterizations**

While a plasmonic reflective filter (PRF) can in principle be created by a variety of methods, here, as a proof-of-concept, we choose what is perhaps the fastest and most convenient way of fabricating one – a galvanic displacement reaction-based dip-and-dry technique, which Mandal et. al. demonstrated for selective solar absorbers.[11] In the particular example shown in Fig. 2A, by dipping a zinc (Zn) strip in aqueous copper sulfate, a PRF comprising of copper (Cu) nanoparticles on zinc is produced. Fig. 2B shows that the nanoparticles, which are clusters with sizes < 200 nm, comprise features < 50 nm in size, indicating that the underlying zinc should play a major role in any TIR reflection. From Fig. 1B and D, it is immediately clear that the PRF is a selective solar absorber and specular TIR reflector. Qualitatively, this serves as proof-of-concept of the PRF's functionality as a reflective filter for thermal imaging systems. We also characterized the optical properties of the PRF by spectrophotometry and interferometry. As shown in Fig. 2C, in this particular embodiment, the PRF has a specular reflectance of 0.007 in the solar and 0.78 in the LWIR wavelengths. The total (diffuse) reflectances are a low 0.028 and high 0.83 respectively in the solar and LWIR, with the non-specular components being 74% and ~ 5% of the total.

The experimental results are consistent with the Mie-scattering behavior of plasmonic nanoparticles in the solar, and effective medium behavior in the TIR wavelengths. The ~5% non-specular reflectance in the LWIR is likely due to surface roughness of the zinc substrate alone, as the copper nanoparticle layer is likely to have a deep-subwavelength thickness.[11] As shown in Fig. 2, both the solar and LWIR non-specular components can be conveniently filtered by adjusting distances between the PRF and the detector or by using apertures. For the purposes of imaging and sensing, therefore, the specular component is of relevance. In that regard, we note that the LWIR:solar yield ratio (i.e. the ratio is reflectances) is a high 110 (~0.78/0.008). This is a useful indicator of the signal to noise ratio, which varies with the intensity of solar and LWIR irradiances incident on the imaging and sensing system.

We also investigated the quality of the optical wavefronts reflected by the PRF, and by extension, its surface quality, by Twyman-Green interferometry at λ~8.1 μm.[12] The upper left panel of Fig. 2D shows the interference pattern observed when the 'test' and 'reference' arms of the interferometer contain a PRF and a standard, vapor-deposited plane aluminum mirror respectively. The pattern is similar to that observed when both arms contain standard mirrors (Fig. 2D, upper right panel), which indicates that the PRF has an optically smooth surface and reflects specularly in the LWIR wavelengths. Additionally, as a proof-of-concept, we also curved a PRF into a concave shape by impressing it with a standard convex mirror. The lower left panel of Fig. 2D shows the interference pattern when the curved mirror is used as the test optic, which is qualitatively similar to the pattern obtained when a concave mirror is used (Fig 2D, lower right panel). Both the interferometric and spectrophotometric characterizations presented here are consistent with the photo- or thermographic evidence (Fig. 1D). Crucially, they also show that the PRF has a high LWIR yield and solar filtration capability – a requisite for IR detection systems – and preserves quality of the reflected wavefronts – which is necessary for imaging. This makes the PRF suitable as both planar and curved optical components for imaging systems (Fig. 2E)

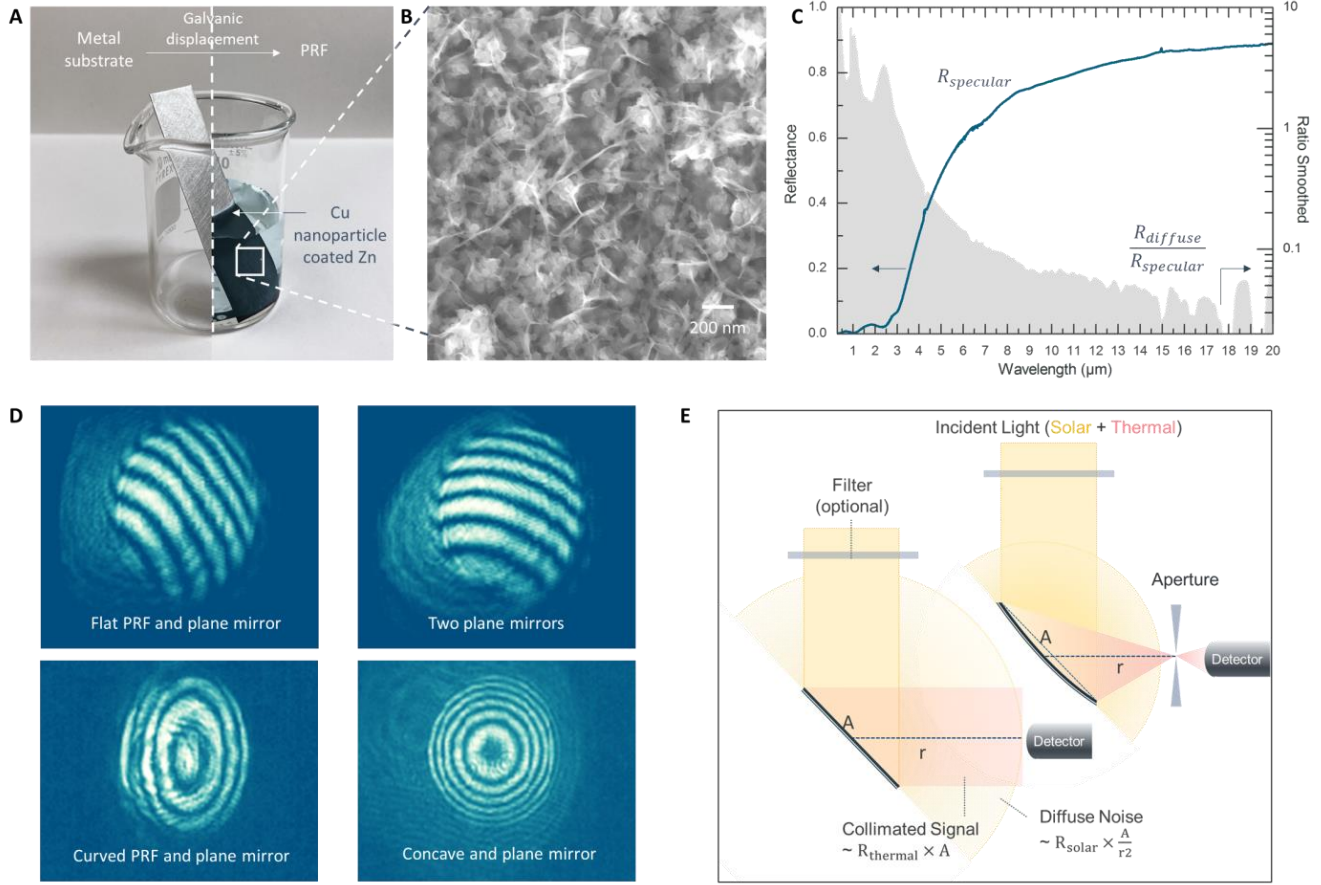

**Figure 2.** Experimental demonstration of a plasmonic reflective filter (PRF). (**A**) Photograph showing a copper nanoparticle coated zinc foil, fabricated using a galvanic displacement reaction-based dip-and-dry technique. (**B**) Scanning electron micrograph showing the copper nanoparticles, with large 200 nm clusters made from smaller nanostructures. (**C**) Specular reflectance of a Cu-Zn PRF shown in blue, with the relative diffuse components shaded in grey. The fluctuation in the longer wavelengths are due to the noise in the diffuse reflectance data. As expected, reflection in the TIR is mostly specular, while that in the solar wavelengths is mostly diffuse. The specular/diffuse solar reflectances are 0.007 / 0.028, and LWIR reflectances are 0.784 / 0.833, respectively. (**D**) (upper left) Twyman green interferogram obtained using a flat PRF and a plane optical grade mirror on the two interferometer arms. The presence of distinct fringes, and the similarities to the case for the two optical grade plane mirrors (upper right), indicates that reflection off the PRF is specular. Analogous interferograms for a roughly curved PRF (lower left) and a plane mirror, and a concave and plane mirror (lower right) show elliptical and circular fringes respectively, and indicate that the PRF could be formed into curved mirrors with focusing functionality as well. (**E**) The results indicate that the PRF can be used as flat and focusing optics in IR sensing and imaging systems. Importantly, the non-specular noise can be filtered out, either by increasing the gap between the PRF and the detector, or using apertures that are common in imaging systems. Additionally, filters can optionally be placed in front of the PRF to impact further optical selectivity.

In addition to optical characterizations, we performed comparative radiometric characterizations of the PRF's imaging capability by comparing direct thermographs of objects with those of their reflections from the PRF. A FLIR Boson Camera (640 x 512 pixel, $\lambda_{range}$ ~7.5-14 µm) was used in the experiments. Fig. 3A shows the variation in temperatures along the dotted lines (insets) across a plastic with silver deposited on one side. We observe that the variations across the sharp boundary between the emissive plastic (bright) and the reflective silver (dark) are very similar, indicating that the insertion of the PRF in the optical path does not impact the sharpness of the image for the given camera resolution. This is promising, since the FLIR Boson 640 Camera's 640 pixel x 512 pixel resolution represents the higher end of thermographic capabilities for typical commercial applications. We also investigated whether the addition of a PRF as an imaging optic raises noise levels in thermographs. This was tested using a thermally stabilized diffuse emitter on a copper block, and taking multiple thermographs of a ~ 1cm x 1cm area of the emitter expected to have a uniform temperature at steady state, directly and off the PRF. Representative images are shown in Fig. 3B. As shown, the direct image of the emitter has mean and standard deviations $26.965 \pm 0.069°C$. The image of the emitter taken off the PRF has mean and standard deviations of $26.224 \pm 0.079°C$. The higher noise levels for the image off the PRF is also observed in the other images taken (bar charts in Fig. 3B) and indicates that addition of the PRF does increase image noise levels by 0.01 K. However, the increase in noise is a fraction (~16%) of the original, and the noise with the PRF is minute compared to temperatures typically measured by thermal cameras (~ 230-420 K). Thus, additional noise arising from the PRF's use is unlikely to be significant.

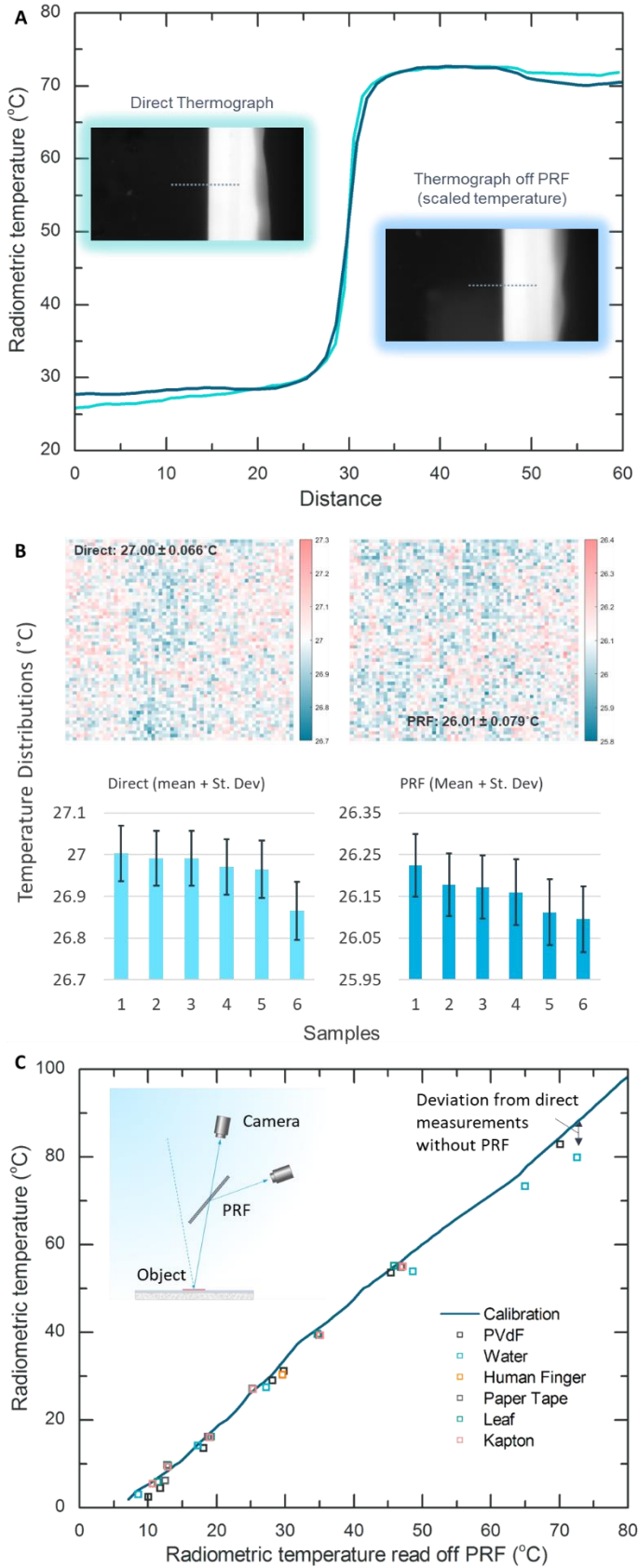

**Figure 3**. Radiometric characterisations involving comparison of thermographs of objects using a high resolution thermal camera, and their reflections off the plasmonic reflective filter (PRF) inserted in the optical path. (**A**) variation in thermographic temperatures along the dotted lines (insets) across a plastic with ~ 200 nm silver deposited on one side, showing that the image sharpness is similar for both the direct and reflected (off PRF) images. The latter was scaled to match the magnitude of the first. (**B**) (Top) Variation in radiometric temperature of an emitter designed to have a spatially uniform temperature, showing fluctuations or noise levels about the mean for both the direct and reflected images. (bottom) Summarized results for multiple direct and reflected images. As shown, the noise levels for images off the PRF is marginally higher, but not significant. (**C**) A preliminary radiometric calibration, performed by taking temperature measurements of an emissive target directly and off the PRF, yields quantitatively close predictions to temperatures measured directly using the thermal camera. Collectively, the results show the integrability of the PRF into thermal imaging and sensing systems without compromising their imaging or radiometric capabilities.

As a further demonstration of the PRF's radiometric capability, we performed a preliminary calibration of the combined thermal camera + PRF system, and used it to measure temperatures of common objects. The calibration was done by taking radiometric temperature measurements of a highly emissive roughened polyvinylidene difluoride film held at different temperatures from -5 to 105°C directly with the thermal camera, and its reflection from the PRF (Fig. 3C, inset). The results were then used to create a calibration curve that predicts the 'real' temperature as would be recorded by the camera alone. The setup was then used to measure temperatures of a range of everyday objects (a leaf, paper, Kapton tape, human skin, a water drop, and smooth PVdF plastic) – both from the PRF and directly with the camera. The directly measured temperatures were then compared to the predictions based on the temperatures recorded using the PRF and the calibration curve.

As can be seen in Fig. 3C, the measured and predicted radiometric temperatures show good agreement. We do note that in our demonstrations, the directly measured temperatures are often slightly (1-3°C) lower than predictions based on the calibration, except for the large differences seen for water at high temperatures. The former is likely due to different ambient radiative temperatures for the calibration and the measurements, which can influence the reflected radiance from the objects being measured and the emitted radiance from the slightly emissive PRF. The latter was due to the heating of the PRF by the water vapor emanating from the hot water below it. These variations are expected, as compared to our rudimentary process, calibrations of radiometric systems are carried out in highly controlled and measured environments, and for multiple variables in addition to temperature. Nonetheless, the closeness of the prediction and the direct measurements indicate that PRFs integrated into thermal sensing systems can be used radiometric measurements as well. This proof-of-concept demonstration thus lays the basis for advanced calibrations (with precisely controlled values of object, ambient, PRF, Ge-lens and camera core temperatures, humidities and shortwave noise levels), which would be required for the PRF's real-life use.

**Comparison with State-of-the-art Thermal Optical Components**

While we have demonstrated the plasmonic reflective filter's (PRF's) suitability for thermal imaging and sensing systems from an optical and radiometric perspective, ultimately, the adoption of this approach will depend on its benefits as well as limitations relative to state-of-the-art Ge-based optical components. We compare in **Fig. 4A** the optical properties of the PRF and Ge-based transmission optics. As shown on the top-left panel, the specific PRF presented in **Figs. 1** and **2** has a similar solar filtering capability as a 5 mm thick Ge optic, with the PRF's absorptance (0.972) and filterable non-specular reflectance (0.021) adding to 0.993 compared to the Ge-optic's 0.986. The PRF's LWIR yield in this specific case is a reflectance of 0.78, higher than germanium's (0.445) but lower than its antireflection-coated variant (0.928). However, the LWIR:solar selectivity of the PRF is higher in this specific case (Fig. 4A, top-right panel), and it is noteworthy that the tunability of its spectral reflectance could be harnessed to enhance both the LWIR yield and the selectivity. Furthermore, the reflectance of the PRF stretches from λ~3 µm into the far-IR wavelengths (λ > 25 µm) (Fig. 4A, bottom left panel). In contrast, germanium's transmittance is high across λ ~2-16 µm, while silicon's transmittance ranges across λ ~1.2-14 µm. The PRF is thus suitable for use over a wider infrared band.

Being a reflective design, the PRF also has characteristic optical advantages and disadvantages relative to transmissive Ge-optics. One limitation is that reflective designs are less compact than transmissive ones. However, the PRF, with its thin nanoparticle layer on metal, is unlikely to suffer from the chromatic aberration that thick transmissive Ge-optics experience. Furthermore, unlike Ge, the PRF's optical dispersion and transmittance do not depend on temperature, and the example we created using the dip-and-dry method can withstand temperatures of up to 200°C in air.[11] This can simplify radiometric considerations, and allow for a wider operating temperature range for thermal imaging and sensing systems.

The plasmonic reflective filter (PRF) is also fabricated by a convenient, low-cost and versatile process. As demonstrated, the galvanic displacement reaction-based dip-and-dry method is a fast and highly convenient way of making PRFs. Given the generalizability of the method, it is compatible with a variety of widely-available and low-cost metals. By comparison, the grinding of brittle Germanium optics[21] the material costs and thickness (~ 5 mm) involved, and the application of antireflection coatings, all add to the cost of Ge-based optical components. Thus, while we made our PRFs at costs of around US$ 1, Ge-based optics cost in the US$ 100-1000 range. The orders of magnitude difference in price persists even with the assumption that production costs are as low as ½ of the selling value.

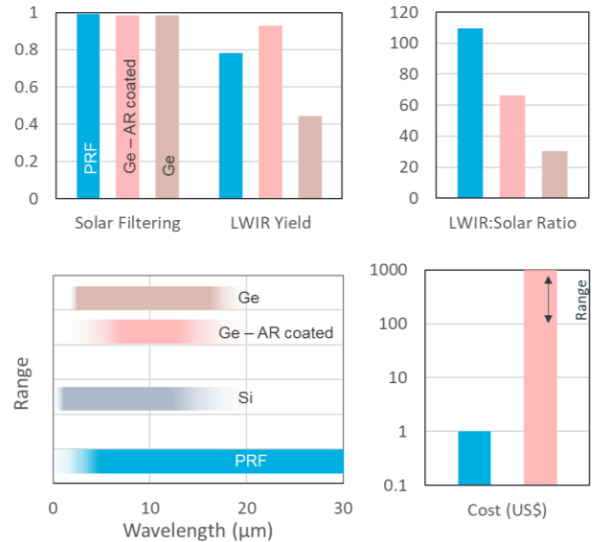

**Figure 5.** Comparisons of the PRF's solar filtering, LWIR yield, ratio of LWIR signal to solar noise, wavelength range and cost to that of Ge optics. In general, the optical properties are comparable, but the PRF is created at orders of magnitude lower costs.

**Outlook**

In this work, we have demonstrated a plasmonic reflective filter that achieves a similar optical performance as state-of-the-art Ge-based optics, but for a fraction of the cost and through a vastly simpler fabrication process. Imaging in the visible and LWIR, spectrophotometry and interferometry corroboratively indicate that the PRF has an efficient LWIR yield, solar filtering capability, and specular reflectance – requisites for thermal imaging and sensing systems. Radiometric characterizations indicate that the PRF can be readily integrated into thermal imaging and sensing systems without compromising their performance. In practical terms, the $10^2$-$10^3$ reduction in cost, and the simplicity and versatility of the fabrication method represents a major advance. The former could be an overriding consideration in terms of use, as IR imaging and sensing expands into diverse, and potentially unforeseen applications. However, we believe that the significance of this work goes beyond immediate practical ones, and could be the basis for further, meaningful explorations.

We first note that while surfaces with selective solar absorptance have been created for solar heating applications, this work is, to our knowledge, the first instance where their potential for imaging and sensing have been explored. As such, this work reports a previously unexplored optical functionality, and highlights the potential of disordered plasmonic media as a platform for the design of infrared optical components. Of perhaps greater significance is the fact that it was achieved using an electrochemical process that is unusual for optical design. We note that a wide array of chemical processes have been used to create plasmonic selective solar absorbers, and could, in practice, be used to create PRFs as well. We therefore envision that this work will spur further exploration of non-traditional chemical fabrication pathways for novel, high-performance and low-cost infrared optical designs.